\documentclass[superscriptaddress,groupedaddress,nofootnoteinbib,12pt]{article}  

\usepackage{amsmath}
\usepackage{amsfonts}
\usepackage{authblk}
\usepackage{bbm}
\usepackage{color}
\usepackage[dvipsnames]{xcolor}
\usepackage{graphicx,graphics}
\usepackage{epstopdf}
\usepackage{caption}
\usepackage{subcaption}
\usepackage{float} 
\usepackage{pifont}
\usepackage{titlesec}
\usepackage{etoolbox}
\usepackage{indentfirst}
\usepackage{jcappub}
\usepackage{braket}

\def\bb{\begin{equation}}
\def\ee{\end{equation}}
\def\ba{\begin{array}}
\def\ea{\end{array}}
\def\babc{\begin{subequations}}
\def\eabc{\end{subequations}}

\def\5{\hspace*{5mm}}
\def\2{{\scriptstyle\frac12}}

\def\dd{\partial \partial}

\begin{document}

\begin{flushright}
NYU-TH-12/13/18
\end{flushright}

\vspace{0.5in}

\begin{center}

\Large{\bf Gravitational Dressing of 3D  Conformal Galileon}

\vspace{0.5in}

\large{Gregory Gabadadze and Giorgi Tukhashvili}

\vspace{0.2in}

\large{\it Center for Cosmology and Particle Physics, Department of Physics, \\
New York University, 726 Broadway,  New York, NY, 10003}

\vspace{0.3in}

\end{center}

Could a conformal  Galileon  be describing  a gauge mode of a  broader diffeomorphism invariant theory?
We  answer this question affirmatively  in  3D   by using a coset construction  for a nonlinearly
realized conformal symmetry. In particular, we show that the conformal Galileon  emerges  as a St\"uckelberg
field of a local Weyl  symmetry. The coset construction  gives us a  diffeomorphism  and Weyl invariant 3D theory containing
first, second, and third  powers of the curvatures, their couplings to certain tensors made of the Galileon and its derivatives, and the conformal
Galileon terms. In a theory with a boundary  additional surface terms are required for the boundary Weyl anomaly cancellation.
When gravity is switched off, the  theory  reduces to the  conformal Galileon plus a boundary term.
On the other hand, it reduces to a generalization of ``New Massive Gravity" by A. Sinha,  if the Weyl
symmetry is gauge-fixed in a particular way. Last but not least,  there exists a parameter space for which the
theory has no propagating  ghosts. Thus we show by an explicit construction that ``New Massive Gravity"
(and its generalization) is a unique scalar ghost free low energy effective field theory non-linearly realizing conformal symmetry in 3D.

\newpage



\section{Introduction and summary}


A Lagrangian for a  free scalar field can be amended by the mass and nonlinear interaction terms  that are polynomial
in the field.  Theories  with up to a quartic polynomial are motivated by the renormalisability in 4D (notwithstanding the
Landau pole), and are often used in  particle physics and cosmology. For a more general polynomial such a theory is not renormalizable but can be quantized as an effective field theory; the quantum loop-induced  interactions with derivatives would  be suppressed by powers of a momentum as compared to the renormalized polynomial terms.

Alternatively, one could think of a single scalar field theory that is made nontrivial by
interaction terms  containing derivatives of the field,  while the polynomials
would not appear due to symmetry reasons (e.g., field space shift symmetry or Galilean invariance).
Generically such theories cannot be truncated  to  a finite   number of interaction terms as this would give rise to a ghost;
instead, they are usually viewed as a low energy approximation to a certain microscopic  theory.
A well known example is the chiral Lagrangian for the Nambu-Goldstone mesons
of massless QCD,  that gets completed into the asymptotically free quark-gluon
Lagrangian;  embedding of General Relativity (GR) into string
theory is another example.

However, there seems to be an exception from the above lore for the higher derivative theories  within
a class  that yields only second
order equations of motion  \cite{Horndeski}. A subset of this class  contains a scalar theory that  emerged
within the brane  induced gravity  \cite {DGP} in a certain limit  \cite {Porrati_gal},  and then
in a more general setting \cite {Rattazzi_gal}, where astrophysical and  cosmological applications were discussed
and the name Galileon introduced.

The Galileon  action exhibits field space galilean invariance  while its   Lagrangian  transforms as a total derivative.
Owing to this property, there are only a  finite number of  Galileon terms in any finite number of dimensions, and they
contain less derivatives per field than a generic  galilean invariant Lagrangian term would.  Because of this, the
generic Lagrangian terms -- that are exactly Galilean invariant --  are suppressed  by powers of derivatives
as compared to the Galileons, and hence one can focus on the latter in a low energy approximation\footnote{A trivial example:
$(\partial \pi)^2$ is a Galileon, while the galilean invariant term, $(\partial \partial \pi)^2$, is not.}. Importantly,
the truncation of the theory to the Galileon terms  does not  lead to additional degrees of freedom.
This is unusual but a perturbative quantum low energy effective field theory with a finite number of Galileons has a
well defined range of validity,  and is predictive \cite {Rattazzi_gal,Rattazzi_eft,Kurt_gal,Lam_gal}.

In spite of this progress in  the perturbative quantum effective field theory, there are issues
in the classical theory related to the lack of global hyperbolicity of the respective nonlinear equations of motion:
while they contain only second derivatives acting on a field, $\dd \pi$,
they are also at least quadratic in $\dd \pi$, leading in general
to issues  of conceptual  and computational nature \cite {Adams_gal}.

Understanding whether or not such issues  represent an impasse  is of
great interest for both  a formal field theory of Galileons and  their potential applications in
cosmology.  The present paper does not aim to study  hyperbolicity for Galileons,
but rather to provide a possibly useful framework for addressing it as follows:
It might be easier to tackle the issue if the Galileon is embedded into a diff invariant theory
of gravity in which it would  just be a gauge mode of a tensor field that enters with no more derivatives than it does
 in GR.  For instance, in both the
brane induced gravity \cite{DGP} and the  diff invariant  massive gravity  \cite {dRG, dRGT},
the Galileon  appears as a gauge mode  describing the helicity-0
state of a massive graviton in a certain limit. Only in that specific limit the Galileon acquires a
gauge-invariant meaning.  Away from the limit  one could remove it by  turning to the unitary gauge, i.e., absorb
it into the metric $g_{\mu\nu}$.  Its classical equations of motion are Einstein's  equations amended
 by a polynomial  in the metric and its inverse, but no extra derivatives. Moreover, the Bianchi identities lead to a constraint that is similar to a harmonic gauge-fixing condition of GR.  Hence, the issue of hyperbolicty in massive gravity in unitary gauge seems to be  at least somewhat similar to that
 in GR\footnote{The results of \cite {Cheung_mg} might be interpreted
 as supporting a thought that the embedding  of the Galileons into a  gravitational
 theory helps with hyperboliciy.}.

Motivated by this, we set a goal to embed the conformal Galileon into a
diffeomorphism invariant theory, where the conformal Galileon would be a gauge mode.
In the present paper this will be done in 3D.  A sequel will address a 4D  construction.

The paper is organized as follows: In section 2 we briefly collect the well known facts about
the coset construction for a conformal group  that has  a Poincar\'e subgroup realized linearly, and the rest nonlinearly.

Section 3 constitutes the bulk of the paper. Using the coset fields we first construct the effective Largangian in the absence
of a dynamical gravitational field; we show that the inverse  Higgs constraint emerges  as a solution to the classical
equations of motion,  and after solving it, reproduce the  Lagrangian  of the conformal Galileon field theory.
{\it En route} we find that the conformal  Galileons can be obtained from a Lagrangian that is identical to that
of the ghost free massive gravity  \cite {dRGT},  but with  a different   St\"uckelberg sector.

Next, we introduce a dynamical gravitational field  by gauging the tangent-space Poincar\'e symmetry.  We then  find the effective Lagrangian.
It contains  four independent sets of terms that  enter with arbitrary coefficients not specified by the coset construction.
The  key point is that  the conformal Galileon  naturally emerges  as a St\"uckelberg
field of a local Weyl  symmetry. The obtained theory is  diffeomorphism  and Weyl invariant.  In addition
to the conventional terms of a Weyl invariant 3D GR,  it contains second and third  powers of the curvatures, their couplings to certain
tensors made of the Galileon and its derivatives, and the full set of the conformal Galileon terms. If a boundary is introduced,
then additional surface terms are required  for the boundary Weyl anomaly  cancellation.
When gravity is switched off, the  theory  reduces to the  conformal Galileon plus a boundary term. On the other hand, it reduces to a generalization of ``New Massive Gravity" \cite {Bergshoeff:2009hq} by A. Sinha \cite {Sinha:2010ai},  if the Weyl symmetry is gauge-fixed in a particular way. Therefore we prove by construction that ``New Massive Gravity" plus its curvature cubed generalization is a unique theory realizing  spontaneously broken conformal symmetry in 3D. Alternatively, one can gauge-fix a conformal  mode of the metric tensor by imposing, e.g.,
unimodularity, $det\,{g}=1$, in which case one would  obtain the covariantized conformal Galileons coupled to unimodular 3D gravity.

\vspace{0.2in}


\section{The $SO(n,2)/ISO(n-1,1)$ coset and conformal Galileons}

We start by a  brief summary of  the coset construction for spacetime symmetries \cite {Salam,Volkov,Ivanov} and
in particular  for a  conformal group spontaneously broken down to the Poincar\'e group \cite{Volkov,Borisov}.
The method  has been  nicely reviewed in \cite{Goon:2012dy} and  was used
to construct the Lagrangians of conformal Galileons in a flat space. In the next section we will generalize the
method to arbitrary gravitational backgrounds by gauging the tangent-space Poincar\'e symmetry (see \cite{Delacretaz:2014oxa} and references therein).

The $n$ dimensional conformal algebra is realized through the $n(n+1)/2$ generators of the Poincar\'e algebra with the generators conventionally denoted by $J_{ab},P_a$, plus $n$ generators of special conformal transformations, $K_a$,  and a generator of dilatations, $D$,  with the well-known  commutation relations:
\begin{align}
\left[ P_a, D \right] = P_a, \5\5\5\5\5 ~~~~~~~& \left[ D, K_a \right] = K_a, \\
\left[ J_{ab}, K_c \right] = \eta_{a c} K_b - \eta_{b c} K_a, \5\5~ & \left[ J_{ab}, P_c \right] = \eta_{a c} P_b - \eta_{b c} P_a , \\
\left[ K_a, P_b \right] = 2 J_{ab}- 2 \eta_{a b} D , \5\5\5 & \left[ J_{ab}, J_{cd} \right] = \eta_{ac} J_{bd} - \eta_{bc} J_{ad} + \eta_{bd} J_{ac}
- \eta_{ad} J_{bc}\,.
\end{align}
Following  the coset construction method  we should look at a  Maurer-Cartan form built out of the $SO(n,2)/ISO(n-1,1)$ coset element $\Sigma$. A convenient parametrization for $\Sigma$ is:
\bb
\Sigma = e^{\pi D} e^{\xi^c K_c}\,.
\ee
Since the Maurer-Cartan form is an element of the Lie algebra we may expand it in the basis of the generators:
\bb\label{maurier_cartan}
\Sigma^{-1} \left ( d  +  dx^a P_a \right ) \Sigma =
E^a P_a + \omega_K^a K_a + \omega_D D - \frac12 \omega_J^{ab} J_{ab}\,,
\ee
where\footnote{Some clarifications: $\delta^a \equiv \delta^a_\mu dx^\mu$. We are distinguishing the Latin indices from
the Greek indices for the later convenience; when we introduce gravity
the former will refer to a Local Lorentz group, while the latter to a space-time group.
$\xi^a$ is a zero form in space-time but a vector of the Local Lorentz group.}:
\bb\label{flat_coset_1}
E^a = e^\pi \delta^a ,
\ee
\bb\label{flat_coset_2}
\omega_D = d \pi + 2 E^a \xi_a,
\ee
\bb\label{flat_coset_3}
\omega_K^a = d \xi^a - \xi^2 E^a + \xi^a \omega_D,
\ee
\bb\label{flat_coset_4}
\omega_J^{ab} = - 2 E^a \xi^b + 2 E^b \xi^a\,.
\ee
Here $\pi$ and $\xi^a$ are the coset fields of  nonlinearly realized dilatations and special conformal transformations,
which may or may not  end up giving rise to the respective dynamical  Nambu-Goldstone fields.  Indeed,
since $\left[ K_a, P_b \right] \sim \eta_{a b} D$, not all the Nambu-Goldstone fields parametrizing the coset
should be physical and it  should be possible
to eliminate the ones associated with the generator $K_a$. This could be achieved by imposing the inverse Higgs constraint (IHC)
$\omega_D = 0$ \cite{Ivanov,Goon:2012dy}.

In the present  paper we will take a different path, instead of imposing the IHC by hand at the level of the algebra
we will  show that it emerges  as a consequence of the classical equation of motion for a properly built action.

\clearpage


\section{The effective Lagrangian}


\subsection{Flat background metric}


Simplest actions in 3D constructed via the coset elements of the last section are \cite{McArthur:2010zm}:
\begin{align}
	S_0 & = \int \varepsilon_{abc} E^a \wedge E^b \wedge E^c ;
	\5\5\5~
	\delta_\xi S _0 = 0 \,, \\
	S_1 & = \int \varepsilon_{abc} E^a \wedge E^b \wedge \omega_K^c ;
	\5\5\5
	\delta_\xi S_1 \simeq - \int \varepsilon_{ab f} ~ \omega_D \wedge E^a \wedge E^b ~ \delta \xi^f\,,\\
	S_2 & = \int \varepsilon_{abc} E^a \wedge \omega_K^b \wedge \omega_K^c ;
	\5\5\5
	\delta_\xi S_2 \simeq 2 \int \varepsilon_{ab f} ~ \omega_D \wedge E^a \wedge \omega_K^b ~ \delta \xi^f \,,\\
	S_3 & = \int \varepsilon_{abc} \omega_K^a \wedge \omega_K^b \wedge \omega_K^c ;
	\5\5\5
	\delta_\xi S_3 \simeq 9 \int \varepsilon_{ab f} ~ \omega_D \wedge \omega_K^a \wedge \omega_K^b ~ \delta \xi^f\,.
\label{action4}
\end{align}
We did not include derivative terms as they'd give  additional  degrees of freedom.
$\omega_D$  has no Latin indices and does not   enable to construct  any nontrivial  coset
term  (see, however,  the section on the Wess-Zumino terms). We've also written out above
the variations with respect to $\xi$ (the sign $\simeq$ denotes
equality up to a total derivative); these expressions
confirm  that IHC, $\omega_D = 0$, is a solution of the equations of motion.

The total action is a sum of the above four actions each weighted with some coefficient which
cannot be determined by the effective action formalism but instead should be constrained
by various  consistency requirements and experimental/observational results whenever available.

Here, we note that  one can rewrite
the above actions in a form equivalent to that of  3D massive gravity potentials \cite {dRGT}
but with a different  matrix $\mathcal{K}$ entering those potentials.
Indeed, define the matrix $\mathcal{K}$  as follows:
\bb
\omega_K^a = E^a_\nu \mathcal{K}_\mu^\nu dx^\mu\,,
\ee
then use IHC, $\omega_D=0$,  to deduce  the expression for $\mathcal{K}$ in terms of $\pi$:
\bb
\mathcal{K}_\mu^\nu = - \frac12 e^{- 2 \pi} \Big( \partial_\mu \partial^\nu \pi -
\partial_\mu \pi \partial^\nu \pi + \frac12 \left( \partial \pi \right)^2 \delta^\nu_\mu \Big)\,,
\ee
and finally, substitute this expression into the above four actions, to get the four conformal  Galileon terms:
\begin{align}
S_0 = & - 6 \int d^3 x ~e^{3 \pi} \,,\\
S_1 = & - 2 \int d^3 x ~e^{3 \pi} \left[ \mathcal{K} \right]
\simeq - \frac12 \int d^3 x ~e^{\pi} \left( \partial \pi \right)^2 \,,\\
S_2 = & - \int d^3 x ~e^{3 \pi} \Big( \left[ \mathcal{K} \right]^2 - \left[ \mathcal{K}^2 \right] \Big)  \simeq - \frac{1}{8} \int d^3 x ~e^{- \pi} \left( \partial \pi \right)^2 L_{1}^{TD}  \,, \\
S_3 = & - \int d^3 x ~e^{3 \pi} \Big( \left[ \mathcal{K} \right]^3 - 3 \left[ \mathcal{K} \right] \left[ \mathcal{K}^2 \right] + 2 \left[ \mathcal{K}^3 \right] \Big) \simeq \\
\nonumber \simeq & \frac{3}{16} \int d^3 x ~e^{-3 \pi} \Big( \left( \partial \pi \right)^2 L_2^{TD} - \frac{3}{2} \left( \partial \pi \right)^4 L_1^{TD} + \left( \partial \pi \right)^6 \Big)\,,
\end{align}
where,  $L_{1,2}^{TD} $ are the total derivative terms of the respective order, $L_{n}^{TD}\sim (\partial \partial \pi)^n$.


\subsection{Dynamical metric}


The goal of this section is to dress the conformal Galileons with a dynamical gravitational field.
In particular, we'd like to arrive at a theory
in which a conformal Galileon field describes a gauge mode of a certain local symmetry. For this, we
will follow a method used in \cite{Delacretaz:2014oxa}; it consists of gauging the Poincar\'e group by  introducing new gauge fields associated with the translations, $e^a$ (vielbein),  and rotations $\omega^{ab}$ (spin connection). The exterior derivative in (\ref{maurier_cartan}) must be replaced with the  exterior covariant derivative. As before,
we use the expansion of the Maurer-Cartan form in the basis of generators:
\bb\label{maurier_cartan_grav}
\Sigma^{-1} \Big( d + e^a P_a - \frac12 \omega^{ab} J_{ab} \Big) \Sigma =
E^a P_a + \omega_K^a K_a + \omega_D D - \frac12 \omega_J^{ab} J_{ab}\,.
\ee
The covariantized counterparts  of (\ref{flat_coset_1})-(\ref{flat_coset_4}) are:
\bb\label{dyn_coset_1}
E^a = e^\pi e^a\,,
\ee
\bb\label{dyn_coset_2}
\omega_D = d \pi + 2 E^a \xi_a\,,
\ee
\bb\label{dyn_coset_3}
\omega_K^a = D \xi^a - \xi^2 E^a + \xi^a \omega_D\,,
\ee
\bb\label{dyn_coset_4}
\omega_J^{ab} = \omega^{ab} - 2 E^a \xi^b + 2 E^b \xi^a\,.
\ee
It should be emphasized that $e^a = e^a_\mu dx^\mu$ is the vielbein of the pure gravitational field, while $D$ is the covariant derivative with respect to $e^a$, and $\omega^{ab}$ is the spin connection of $e^a$. Using $\omega_J^{ab}$ we can build another coset element - the Riemann curvature two form:
\bb
\mathcal{R}^{ab} = d \omega_J^{ab} + {{\omega_J}^{a}}_c \wedge {\omega_J}^{cb} = R^{ab} + 2 E^a \wedge \omega_K^b + 2 \omega_K^a \wedge E^b\,,
\ee
where
\bb
R^{ab} = d \omega^{ab} + {{\omega}^{a}}_c \wedge {\omega}^{cb}\,.
\ee
The curvature two form   can  be used, alongside with the  fields utilized in the last section,
to construct the effective Lagrangian terms.

Before proceeding to this task, we recall that  a gravitationally covariantized conformal unitary field theory must be Weyl invariant \cite{Farnsworth:2017tbz}. To incorporate this property, we define the Weyl transformations as:
\bb
e^a \rightarrow e^{\sigma} e^a , \5\5 \pi \rightarrow \pi - \sigma , \5\5 \xi^a \rightarrow \xi^a + \frac12 e^{-\pi} \partial^a \sigma\,.
\ee
It is easy to verify that $E^a, ~\omega_D$, and $\mathcal{R}^{ab}$ are all
invariant under these transformations. Moreover,
since $\omega_K^a$ is part of $\mathcal{R}^{ab}$ it would be logical to construct the
effective Weyl invariant actions using the above three elements only.

In $3D$ the Weyl tensor is identically zero and $R^{ab}$ is entirely determined in terms of the Ricci tensor and scalar,
so it is more convenient to work in terms of these quantities rather than the Riemann two form $R^{ab}$ itself:
\bb\label{curv_2_3d}
R^{ab} = e^a \wedge {\bar S}^b + {\bar S}^a \wedge e^b \,, \5\5\5 {\bar S}^a = R^a - \frac{1}{4} R e^a\,,
\ee
\bb\label{weyl_curv_3d}
\mathcal{R}^{ab} = E^a \wedge \Omega^b + \Omega^a \wedge E^b\,, \5\5\5 \Omega^a = \frac{1}{2} e^{-\pi} {\bar S}^a + \omega_K^a\,.
\ee
Here $R^a \equiv e^a_\nu R^\nu_\mu dx^\mu$ and $R$ are the Ricci one form and scalar respectively. ${\bar S}^a$ is the first order analogue of the Schouten tensor, we will refer it as a Schouten one form. This quantity has interesting properties in $3D$ which are inherited from the curvature two form. In particular, from (\ref{curv_2_3d}) we see that it satisfies both Bianchi identities:
\bb\label{bianchi_3d}
D {\bar S}^a = 0 , \5\5\5 e_a \wedge {\bar S}^a = 0\,.
\ee
Expressions (\ref{curv_2_3d}) and (\ref{bianchi_3d}) enable  us to conclude that, in $3D$, ${\bar S}^a$ fully replaces the curvature two form, $R^{ab}$. Following the same logic, from (\ref{weyl_curv_3d}) we also see that $\Omega^a $ can replace $\mathcal{R}^{ab}$ as a coset element. Thus there are three Weyl invariant coset elements that can be used to build effective actions: $\omega_D, ~E^a$ and $\Omega^a$. The rest is straightforward, building the actions follows the same recepe as in the case of the flat metric. Actions and their variations with respect to $\xi$ are given by:
\begin{align}
\mathcal{S}_0 & = \int \varepsilon_{abc} E^a \wedge E^b \wedge E^c ;
\5\5\5
\delta_\xi \mathcal{S}_0 = 0 \,, \label{cov_act_0} \\
\mathcal{S}_1 & = \int \varepsilon_{abc} E^a \wedge E^b \wedge \Omega^c ;
\5\5\5
\delta_\xi \mathcal{S}_1 \simeq - \int \varepsilon_{ab f} ~ \omega_D \wedge E^a \wedge E^b ~ \delta \xi^f \,,  \label{cov_act_1}\\
\mathcal{S}_2 & = \int \varepsilon_{abc} E^a \wedge \Omega^b \wedge \Omega^c ;
\5\5\5
\delta_\xi \mathcal{S}_2 \simeq 2 \int \varepsilon_{ab f} ~ \omega_D \wedge E^a \wedge \Omega^b ~
\delta \xi^f \,,  \label{cov_act_2}\\
\mathcal{S}_3 & = \int \varepsilon_{abc} \Omega^a \wedge \Omega^b \wedge \Omega^c ;
\5\5\5
\delta_\xi \mathcal{S}_3 \simeq 9 \int \varepsilon_{ab f} ~ \omega_D \wedge \Omega^a \wedge \Omega^b ~ \delta \xi^f \,. \label {cov_act_3}
\end{align}
The actions, $\mathcal{S}_0, ~\mathcal{S}_1, ~\mathcal{S}_2$, and $\mathcal{S}_3$,  are respectively: the  Weyl invariant analogues of the cosmological term, the   Einstein-Hilbert kinetic term, the  Bergshoeff, Hohm \& Townsend term \cite{Bergshoeff:2009hq} (usually referred as ``New Massive Gravity"),  and the curvature cubed term \cite {Sinha:2010ai}. It is clear  that IHC remains to be a solution of the equations of motion. Substituting that solution,
the actions  can be rewritten  in the following  form:
\bb
\mathcal{S}_0 = - 6 \int d^3 x \sqrt{g} e^{3 \pi} \,,
\ee
\bb
\mathcal{S}_1 = -2 \int d^3 x \sqrt{g} e^{\pi} \Big( \frac{1}{8} R + e^{2 \pi} \left[ \mathcal{K} \right] \Big)\,,
\ee
\begin{align}
\nonumber \mathcal{S}_2 & = - \int d^3 x \sqrt{g} e^{- \pi} \Big[ - \frac{1}{4} \Big( R_{\mu \nu} R^{\mu \nu} - \frac{3}{8} R^2 \Big) - e^{2 \pi} \Big( R^\mu_\nu - \frac12 \delta^\mu_\nu R \Big) \mathcal{K}_\mu^\nu + \\
{} & ~~~~~~~~~~~~~~~~~~~~~~~~~ + e^{4 \pi} \Big( \left[ \mathcal{K} \right]^2 - \left[ \mathcal{K}^2 \right] \Big) \Big]\,,
\\
\nonumber \mathcal{S}_3 &  = - \int d^3 x \sqrt{g} e^{- 3 \pi} \Big[ \frac{1}{4} R_\mu^\nu R_\nu^\rho R_\rho^\mu - \frac{9}{32} R R_{\mu \nu} R^{\mu \nu} +\frac{17}{256} R^3 + \\
\nonumber {} & ~~~~~~~~~~~~~~~~~~~~~~~~~  + \frac{3}{4} e^{2 \pi} \Big( \frac{5}{8} R^2 \delta^\mu_\nu - \frac{3}{2} R R^\mu_\nu - R_{\alpha \beta} R^{\alpha \beta} \delta^\mu_\nu +2 R^\mu_\rho R^\rho_\nu \Big) \mathcal{K}_\mu^\nu + \\
{} & ~~~~~~~~~~~~~~~~~~~~~~~~~ + 3 e^{4 \pi} \Big( R^\mu_\nu - \frac{3}{8} R \delta^\mu_\nu \Big) \Big( \mathcal{K}_\mu^\rho \mathcal{K}_\rho^\nu - \left[ \mathcal{K} \right] \mathcal{K}_\mu^\nu \Big) + \\
\nonumber {} & ~~~~~~~~~~~~~~~~~~~~~~~~~ + e^{6 \pi} \Big( \left[ \mathcal{K} \right]^3 - 3 \left[ \mathcal{K} \right] \left[ \mathcal{K}^2 \right] + 2 \left[ \mathcal{K}^3 \right] \Big) \Big]\,,
\end{align}
with
\bb
\mathcal{K}_\mu^\nu = - \frac12 e^{- 2 \pi} \Big( \nabla_\mu \nabla^\nu \pi - \partial_\mu \pi \partial^\nu \pi + \frac12 \left( \partial \pi \right)^2 \delta^\nu_\mu \Big) \,.
\ee
Thus we've arrived at the action in which  the conformal galileon is  a  Stuckelberg field for the local Weyl invariance. It is also clear that in the limit when
gravity is switched off  $h_{\mu \nu} \rightarrow 0$ (where $h_{\mu \nu} = g_{\mu \nu} - \eta_{\mu \nu}$; $g_{\mu \nu} = e^a_\mu e^b_\nu \eta_{a b}$ being the dynamical metric) the actions $\mathcal{S}_i, i=0,1,2,3$ reduce to their flat space counterparts $S_n,n=0,1,2,3$.

\subsection{Wess-Zumino terms and anomalies}


It is known that there are no conformal anomalies in odd dimensions \cite{Deser:1993yx}.
We can verify this in the case at hand  by explicitly building the Wess-Zumino terms.
This is done by invoking  $\omega_D$, which does not have a Lorentz index and can be used
to create a four form in 3D. From the expressions below we see that the WZ terms
coincide with (\ref{cov_act_0})-(\ref{cov_act_3}) and therefore do not contain any new information:
\begin{align}
	3 \int \varepsilon_{abc} ~\omega_D \wedge E^a \wedge E^b \wedge E^c & =  \mathcal{S}_0 \,, \\
	\int \varepsilon_{abc} ~\omega_D \wedge E^a \wedge E^b \wedge \Omega^c & =  \mathcal{S}_1 \,, \\
	-\int \varepsilon_{abc} ~\omega_D \wedge E^a \wedge \Omega^b \wedge \Omega^c & =  \mathcal{S}_2 \,, \\
	-3 \int \varepsilon_{abc} ~\omega_D \wedge \Omega^a \wedge \Omega^b \wedge \Omega^c & =  \mathcal{S}_3\,.
\end{align}
The absence of anomalies allows us to  eliminate the gauge mode $\pi$ from the actions, therefore the obtained  theory in the unitary gauge coincides with ``New Massive Gravity" \cite{Bergshoeff:2009hq} and its extensions \cite{Sinha:2010ai}.

What happens if we include a 2D boundary?  In that case and for classically scale invariant
degrees of freedom there is a quantum trace anomaly on
the boundary \cite{Graham:1999pm,Solodukhin:2015eca}:
\bb
\int d^3 x \sqrt{g} T^\mu_\mu = \int d^3 x \sqrt{g} \delta (x_\perp) \Big( c_1 r + c_2 k^\mu_\nu k^\nu_\mu \Big)\,,
\ee
where the coefficients $c_1$  and $c_2$ depend on a field content and boundary conditions considered;
 $\delta (x_\perp)$ has support only on the boundary, $r$ and $k^\mu_\nu$ are its intrinsic and traceless part of extrinsic curvature respectively.
Under the Weyl transformation $r$ transforms, while the $\sqrt{g} k^\mu_\nu k^\nu_\mu $ is invariant. The effective action that reproduces the trace
anomaly can be expressed as \cite{Komargodski:2011xv,Jensen:2015swa}:
\bb\label{anomaly}
\mathcal{S}_{\text{anomaly}} = \int d^3 x \sqrt{g} \delta (x_\perp) \Big[ c_1 \left( \pi r +  \left(\partial \pi \right)^2 \right)+ c_2 \pi k^\mu_\nu k^\nu_\mu \Big]\,.
\ee
This action should be included, with the opposite sign,  for  the Weyl symmetry to remain exact at the quantum level. If so then
the  $\pi$  mode can  be eliminated by a gauge fixing. Alternatively,
one could choose to  gauge fix a conformal mode of the metric tensor, by imposing,  e.g.,  a unimodularity condition,
$\det {g}=1$, and thus obtaining unimodular gravity coupled to covariantised conformal Galileons. The boundary terms would be relevant if one were to study the holographic dual of the model in AdS space. Investigations in this interesting
direction are beyond the scope of the present work.


\section{Degrees of freedom}

Last but not least, important comments concerning the degrees of freedom are in order here.
The total  bulk action
\bb
\mathcal{S}_{tot} = \sum_{i=0}^3 \alpha_i \,\mathcal{S}_i\,,
\label{Stot}
\ee
is a linear combination of the four actions with arbitrary coefficients denoted by $\alpha_i$.
These coefficients cannot be calculated  within the low energy effective field theory approach.
However, various consistency conditions can be imposed on them.

Since (\ref {Stot}) in general propagates ghosts,  we  impose the condition that any
type of ghost to be absent  from the theory. Note that these ghosts are neither due to the
conformal Galileons, nor the nonlinear terms of a conformal
mode of the 3D tensor field  \cite {Itay}, which are both ghost free. Instead,   the ghosts
arise because  of the tensor in the  terms quadratic and cubic in the curvatures.

To reveal the ghosts let us begin with  small perturbations above the Minkowski space in the unitary gauge
$\pi=0$.  In that case the graviton propagator received contributions from  $\alpha_1 \,\mathcal{S}_1$ and
$\alpha_2 \,\mathcal{S}_2$, while the cubic curvature terms  give rise to the interactions but not to a modification
of the graviton propagator. Schematically, the graviton propagator is proportional to
\bb
{1\over \alpha_1 \partial^2 + \alpha_2 \partial^4}\,,
\label{prop}
\ee
where  we have ignored the tensorial structure for simplicity. The second derivative
arises from $\mathcal{S}_1 $, while the quartic derivative  comes form $\mathcal{S}_2$.

 The above expression has two poles, describing  two states,  one massless and one massive, as it can
 also be seen from the equivalent rewriting of  (\ref {prop}):
 \bb
{1\over \alpha_1 \partial^2}   - {1\over  \alpha_1 (\partial^2 + \alpha_1/\alpha_2)}\,.
\label{prop1}
\ee
The latter shows  that either  the first  or the second pole has to describe a ghost.  The situation seems hopeless, except
that in 3D a massless tensor does not propagate any dynamical degrees of freedom, hence one  could
afford it to have a  ghost-like sign of the kinetic term. Thus, we choose the coefficients $\alpha_1$ and $\alpha_2$  as in New Massive Gravity,  $\alpha_1=-1$, $\alpha_2>0$ \cite {Bergshoeff:2009hq}. This guarantees that the massive
tensor mode with two dynamical degrees of freedom  has a  good kinetic term and non-tachyonic mass;
it also guarantees that the Galileon has a ``right-sign" kinetic term.

To reiterate,  the above choice corresponds
to a ``wrong"  sign of the Einstien-Hilbert term in the action (\ref {Stot}); as a result,
the massive tensor   mode has good kinetic and mass terms,  while the massless one
has a ghost-like kinetic term. Since a massless
tensor field in 3D does not propagate any dynamical degrees of freedom, such a ghost-like term would not
lead to the usual uncontrollable instabilities; it would  instead manifest itself  as ``antigravity" for
certain classical solutions in 3D \cite {Gaston}.

Regarding the $ \alpha_3 \,\mathcal{S}_3$ term, as mentioned before it gives new interactions on Minkowski space, however
would yield  further four-derivative modifications of the propagator on  a generic curved background;  hence we  will
also put $\alpha_3=0$ to avoid possible ghosts on generic curved backgrounds.

With the above choice of the coefficients the  theory
is just a Weyl invariant generalization  of New Massive Gravity, and could have been
obtained from the latter by a Weyl transformation of the metric field. It is interesting that
this very theory is derived via the coset construction  for a nonlinearly realized conformal
symmetry.  Moreover, it  can be rewritten  as a theory  with  two
tensor fields, one being dynamical and another algebraically determined \cite {Bergshoeff:2009hq} so that only
linear terms in curvature appear in the action; the issue of hyperbolicity that we alluded to in the first section might
be easier to study in this formulation than it is for conformal Galileons.

Alternatively, the theory provides a covariant formulation of the conformal Galileon  in which the conformal mode of
the tensor field can be gauged away instead of the Galileon. In that case, the dynamics of the gravitational
conformal mode is entirely encoded in the conformal Galileon.

If  a boundary is present, then the boundary terms should be introduced
as in (\ref {anomaly}) to maintain the Weyl invariance of the full theory at the quantum level.

\subsection*{Acknowledgements}

We'd like to thank David Pirtskhalava and Siqing Yu for helpful comments.
The work of GG was supported in part by NSF grant PHY-1620039. GT was supported by the NYU James Arthur Graduate Fellowship.



\end{document}